\documentclass[referee]{raa}            

\usepackage{graphicx,times}             
\usepackage{natbib}
\usepackage{amssymb,amsmath}
\usepackage{color}
\bibpunct{(}{)}{;}{a}{}{,}

\usepackage[a4paper=true,pagebackref=true]{hyperref}
\hypersetup{colorlinks = true, linkcolor = green, anchorcolor = red, citecolor = blue, filecolor = red, pagecolor = red, urlcolor = red}

\begin{document}

   \title{The NTSC VLBI System and its application in UT1 measurement
}

   \volnopage{Vol.0 (20xx) No.0, 000--000}      
   \setcounter{page}{1}          

   \author{Dang Yao
      \inst{1,2}
   \and Yuan-Wei Wu
      \inst{1}
   \and Bo Zhang
      \inst{3}
   \and Jing Sun
      \inst{4}
   \and Yan Sun
      \inst{2,3}
   \and Shuang-Jing Xu
      \inst{3}
   \and Jia Liu
      \inst{1}
   \and Lang-Ming Ma
      \inst{1}
   \and Jian-Jun Gong
      \inst{1}
   \and Ying Yang
      \inst{1}
   \and Xu-Hai Yang
      \inst{1}
   }

   \institute{National Time Service Center, Chinese Academy of Sciences, Xi'an, 710600, China; {\it yaodang@ntsc.ac.cn, yuanwei.wu@ntsc.ac.cn}\\
        \and
             University of Chinese Academy of Sciences, Beijing, 100039, China\\
        \and
             Shanghai Astronomical Observatory, Chinese Academy of Sciences, Shanghai 200030, China\\
		\and
			 National Astronomical Observatories, Chinese Academy of Sciences, Beijing, 100101, China\\
\vs\no
   {\small Received~~20xx month day; accepted~~20xx~~month day}}

\abstract{ In order to measure the Universal Time (UT1) in real time, National Time Service Center (NTSC) has built a VGOS-like (VLBI Global Observing System) broadband VLBI network, which includes three 13-m radio telescopes located in Jilin, Sanya and Kashi, and a data analysis center in Xi'an. Each station is equipped with a highly stable hydrogen atomic clock and a self-developed VLBI backend, and is co-located with two GPS receivers. This VGOS-like VLBI network may play an important role in improving the Chinese broadband VLBI technology and making valuable contributions to domestic VLBI measurements of UT1. In this paper, we introduce the specifications of this VLBI network, and present the UT1 measurements at C-band conducted in 2018 using the Jilin-Kashi baseline of this network. The comparisons between our UT1 estimates and those provided by IERS suggest that the NTSC VLBI network is capable to determine UT1 accurate at the level of 58.8 microseconds.
}

   \authorrunning{D. Yao, Y.-W. Wu \& B. Zhang et al.}            
   \titlerunning{The NTSC VLBI System and its application in UT1 measurement}  

   \maketitle

%
%
\section{Introduction}           
\label{sect:intro}

Earth orientation parameters (EOP), including Universal Time (UT1), Polar motion (PMX, PMY), and corrections of the conventional Precession/Nutation model ($d\psi$, $d\varepsilon$), are a collection of parameters that provide the rotation between the International Terrestrial Reference Frame (ITRF) and the International Celestial Reference Frame (ICRF). Thus, EOP are widely used in Astronomy, Geosciences, Navigation, Deep Space Tracking and Precise Orbit Determination. Among the five parameters of EOP, UT1, the measurement of rotation angle of the Earth, is the most unpredictable parameter that irregularly changes day by day, therefore needs daily observations. The International VLBI Service for Geodesy and Astrometry (IVS) organizes two modes of observations to measure the EOP. The regular mode (R1, R4) is 24h experiment with $\sim$10 stations, observed twice a week with UT1 accuracy of about 7 microseconds and a latency of 1-2 weeks (\citealt{Schuh+etal+2010}). The intensive mode (INT1, INT2 and INT3) observes one hour per day using 2-4 stations, with a typical accuracy of about 15 microseconds and a latency less than one day (\citealt{Bohm+etal+2011}).

In order to achieve a better accuracy, continuous data flow, and short latency of EOP products, in 2000s, IVS formed the VLBI2010 Committee and the Implementation Team to define and design the next-generation VLBI system, i.e., the VLBI2010 system. The next-generation geodetic VLBI network is finally named as the VLBI Global Observing System (VGOS). It is characterized by a high temporal resolution of the observations by using fast slewing antennas and a broad radio frequency bandwidth, with the aim of achieving the accuracy of 1 mm station position and 0.1 mm/year velocity on global scales, continuous observing (7 days per week and 24 hours per day), and posting of initial geodetic results within 24 hours after observations are completed (\citealt{Niel+etal+2005}). As of 2018, there are 22 VGOS projects with 24 new radio telescopes already built or under construction.

\begin{figure}[h]
\centering
\includegraphics[width=0.6\textwidth]{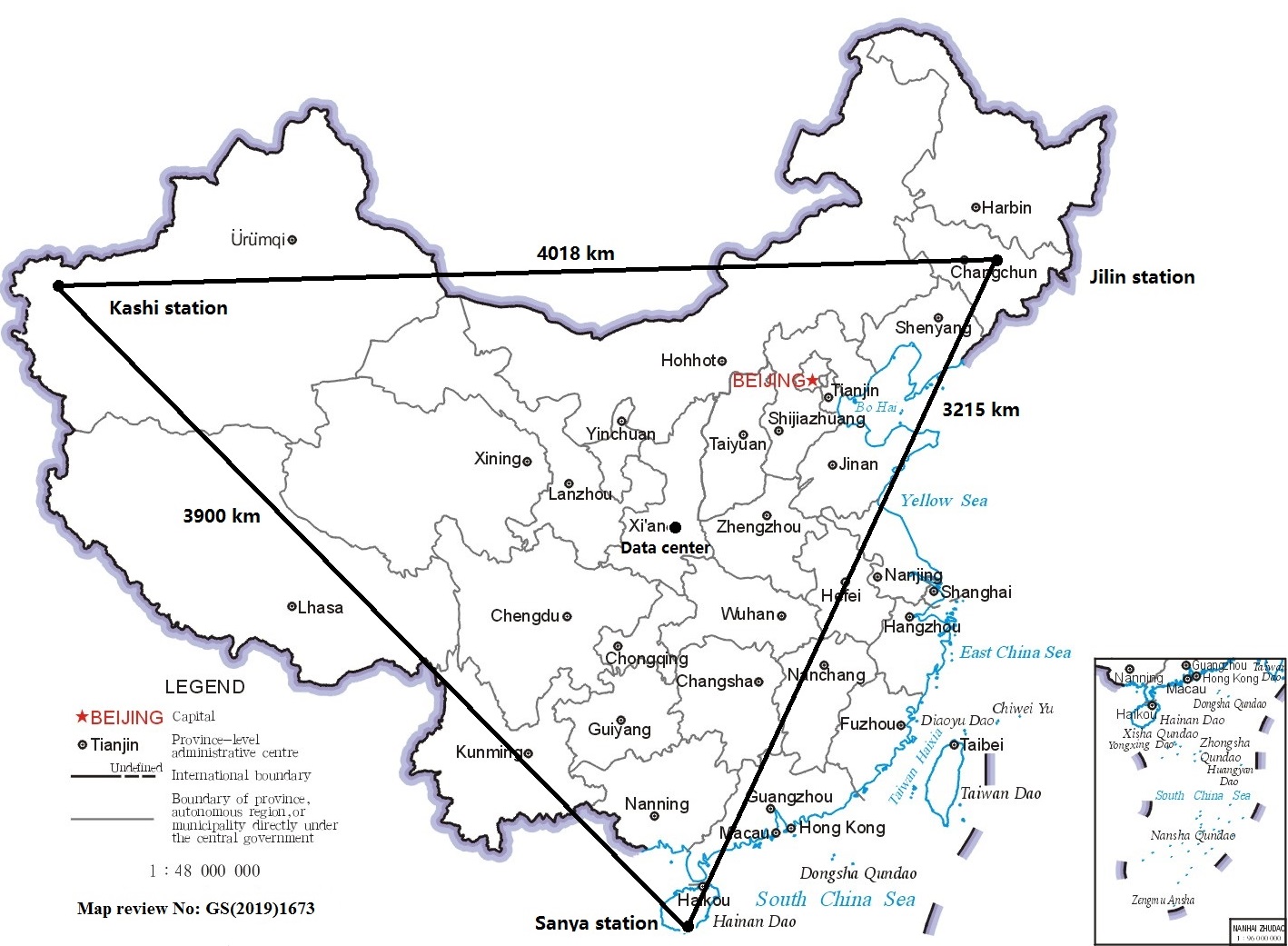}
\caption{The geographical distribution of the NTSC VLBI network, consists of three stations located in Jilin, Sanya and Kashi. The data analysis center is in Xi'an.} 
\label{Fig1}
\end{figure}

National Time Service Center (NTSC)\footnote{NTSC is the National institute that is responsible for producing, keeping and transmitting the national standard time.} of Chinese Academy of Sciences (CAS) started to build a VGOS-like broadband VLBI system since 2013. It consists of three 13-m diameter telescopes with baseline lengths ranging from 3000 km to 4000 km (Figure~\ref{Fig1}) and a data analysis center. The construction was finished in 2016. Following a period of commissioning and debugging, the first fringes for the radio source 3C84 and an artificial GEO (Geostationary Earth Orbit) satellite were successfully detected in the middle of 2016. Nowadays this VLBI system has been used for artificial satellite orbit determination (\citealt{Gong+etal+2018}) and UT1 measurement.

In this paper, we introduce the NTSC VLBI system in Section~\ref{sec:2}, and UT1 experiments in the intensive mode conducted in 2018 are presented in Section~\ref{sec:3}. Analysis and results of the experiments are shown in Section~\ref{sec:4}. Finally, the conclusion and future plans are summarized in Section~\ref{sec:5}. 

\section{NTSC VLBI System}\label{sec:2}
\subsection{Imaging capability}
The geographic coordinates of Jilin, Kashi and Sanya stations are list in Tabel~\ref{tab1}. It is noted that the Sanya station has the ability to observe most of the southern sky sources ($\sigma\geq-70^\circ$). The Jilin-Kashi baseline, which is the longest east-west baseline, may play an important role in domestic UT1 measurement.

\begin{table}[h]
\begin{center}
\caption[]{The Geographic Coordinate of the three NTSC VLBI antennas}\label{tab1}
\begin{tabular}{ccccc}
\noalign{\smallskip}\hline
  Site & Station code & Longitude& Latitude& Above sea level \\
   \hline\noalign{\smallskip}
  Jilin station& JL& $126^\circ$& $43^\circ$& 313 m\\
  Kashi station& KS& $76^\circ$& $39^\circ$& 1242 m\\
  Sanya station& SY& $109^\circ$& $18^\circ$& 12 m\\
\noalign{\smallskip}\hline
\end{tabular}
\end{center}
\end{table}

\begin{figure}[h]
\centering
\includegraphics[width=0.8\textwidth]{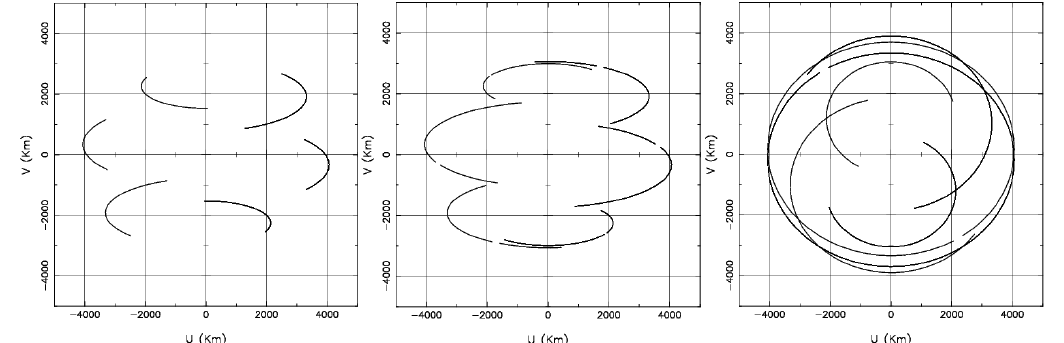}\\
\includegraphics[width=0.8\textwidth]{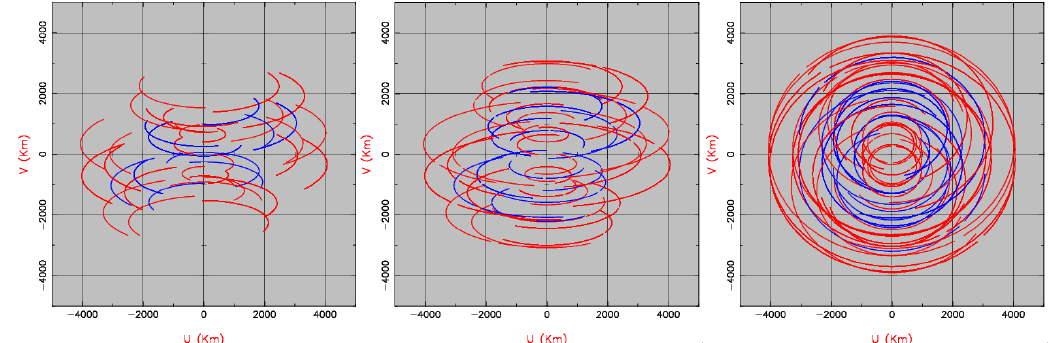}
\caption{The top panels are the simulated u-v coverage with the NTSC VLBI array for the source at declination: -20$^\circ$ (left), 20$^\circ$ (middle), and 60$^\circ$ (right), respectively. The bottom panels are the Simulated u-v coverage with the seven Chinese VLBI antennas for the source at declination: -20$^\circ$ (left), 20$^\circ$ (middle), and 60$^\circ$ (right), respectively.} 
\label{fig2}
\end{figure}

The NTSC VLBI array can also carry out astrometric and astrophysical observations which require properly calibrated visibility amplitudes and phases. In the top panels of Figure~\ref{fig2}, we show the simulated u-v coverage of the sources at different declinations. The NTSC network has a common view of $\sim$7 hours for a source at $Dec = -20^\circ$, $\sim$11 hours for a source at $Dec = 20^\circ$, and $\sim$16 hours for a source at $Dec = 60^\circ$. It can be seen that the u-v coverage of the NTSC VLBI network is sparse, especially, there is a hole due to the lacking of short baseline, which  means a limited imaging capability.

In the bottom panels of Figure~\ref{fig2}, we show the simulated u-v coverage by combing the Chinese VLBI Network (CVN,  \citealt{Zheng+2015})  and the NTSC array, both of them cover the S/X band. The blue tracks are the baselines among KUNMING, URUMQI, MIYUN and SESHAN25, the red u-v tracks are the baselines increased by the three NTSC stations. It can be seen that the longest baseline is increased from $\sim$3000~km to $\sim$4000~km, this means an improvement of the synthetic beam by 30\%. Meanwhile, the combination of the two networks also increases the short baselines, which is essential for study of sources with extend structure, for example the radio jet of M87 (\citealt{Walker+etal+2018}).  

\subsection{Antenna and Receiver System}
\begin{figure}[h]
\centering
\includegraphics[width=0.8\textwidth]{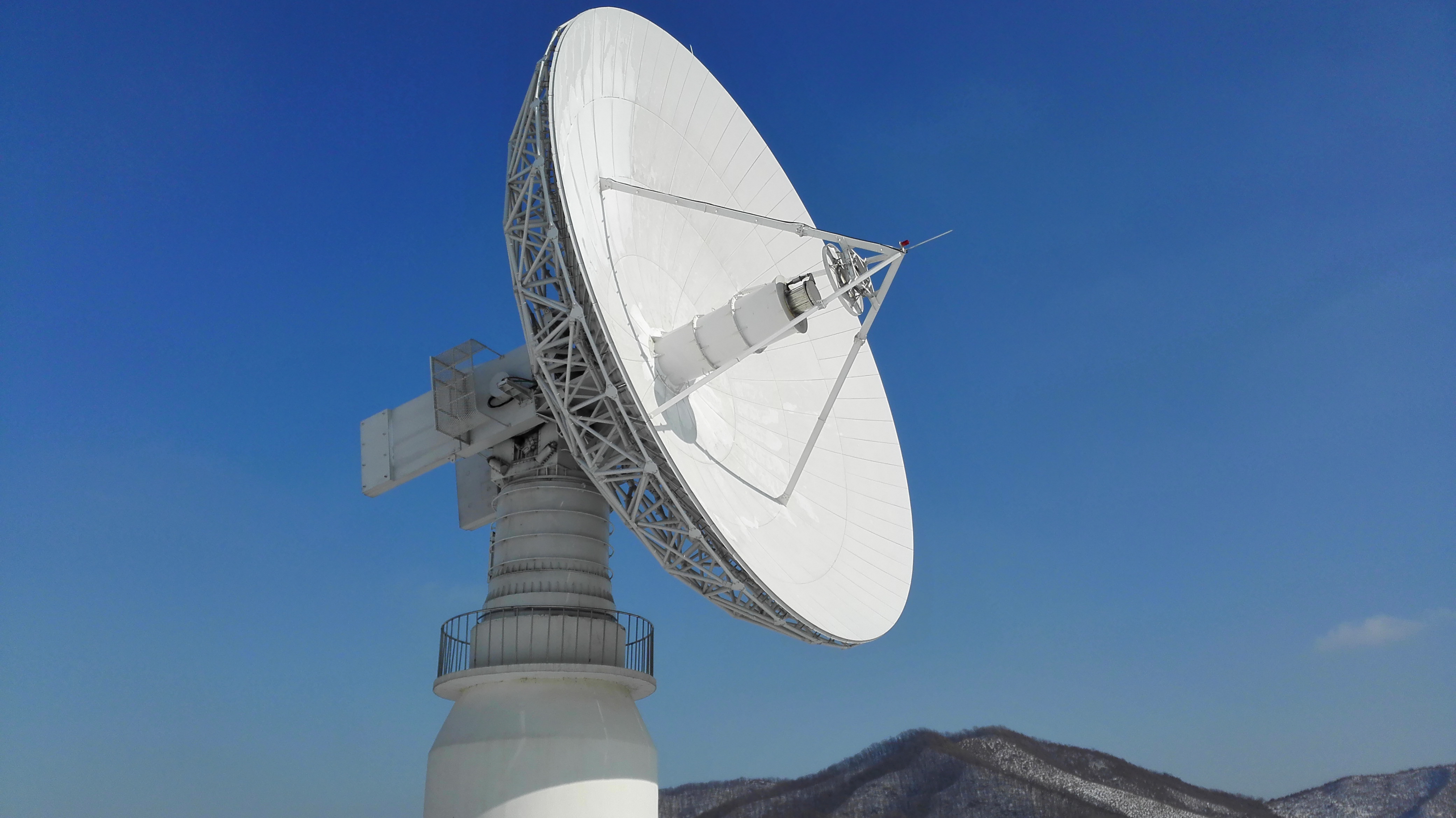}
\caption{The 13-m Cassegrain radio telescope at Jilin station.} 
\label{fig4}
\end{figure}

The three 13-m antennas were designed and constructed by the $39^{th}$ Institute of China Electronics Technology Group Corporation (CETC). The quasi-parabolic mirror of the radio telescope has a diameter of 13 m and stand $\sim$17 meters above the ground. The secondary quasi-hyperbolic mirror was fixed by the quadrapod support structure with diameter of 1.48 m (see Figure~\ref{fig4}). 

The maximal slewing rate of antenna is $12^\circ$/s in azimuth and $6^\circ$/s in elevation, and the minimal slewing rate is $0.002^\circ$/s, which makes it be able to track not only radio sources but also near-earth satellites. The antenna can clockwise and counterclockwise rotate $270^\circ$ relative to the south direction in azimuth and from $5^\circ$ to $90^\circ$ in elevation. By observations of ICRF radio sources (with position accuracy of $\sim$1~mas) and building a 18-parameter modified model, the pointing accuracy is achieved at the level of 30 arcseconds (\citealt{Wu+etal+2018}). The detail information of the antenna system is summarized in Tabel~\ref{tab2}.

\begin{table}[h]
\begin{center}
\caption[]{Specifications of NTSC VLBI antennas}\label{tab2}
\begin{tabular}{ll}
\noalign{\smallskip}\hline
Parameter& Value \\
\hline\noalign{\smallskip}
Antenna mount& Altazimuth \\
Reflector configuration& Cassegrain \\
Main dish Diameter& 13 m paraboloid \\
Main dish Surface accuracy& 0.3 mm rms \\
Secondary refl. Diameter& 1.48 m \\
Secondary refl. Surface accuracy& 0.1 mm rms \\
Pointing accuracy& 30 as \\
Frequency range& 1.2$ \sim $9 GHz \\
Azimuth range$^a$& $-270^\circ \sim +270^\circ$ \\
Elevation range& $5^\circ \sim 92^\circ$ \\
Azimuth slewing rate& $12^\circ$/s \\
Elevation slewing rate& $6^\circ$/s \\
Azimuth/Elevation slewing accel.& $3^\circ$/$s^2$ \\
Minimal slewing rate& $0.002^\circ$/s \\
\hline
$^a$The south is $0^\circ$ \\
\noalign{\smallskip}
\end{tabular}
\end{center}
\end{table}

Cryogenic Cooling broad-band receivers were designed and produced by the $16^{th}$ Institute of CECT. The frequency ranges from 1.2 to 9 GHz for Jilin and Sanya. For Kashi station, due to strong, up to 200 kilowatt, radio frequency interference (RFI) from a nearby meteorologic radar at 5.43 GHz, the frequency band over 5 GHz are temporary filtered out. Thus, Kashi station is currently only available  from 1.2 to 5 GHz. The receivers can receive right- and left-hand circular polarisation  (RCP and LCP) signals simultaneously.

The cryogenic system cools the low noise amplify (LNA) down to $\sim$10-20 K, the noise temperature of the receiver  is about 50 K. Howerver, the typical system temperature measured at 5 GHz towards the zenith direction are 70 K, 75 K and 65 K for Sanya, Kashi and Jilin stations, respectively, due to the various atmospheric noise, ground radiation noise, RFI, etc.

The sensitivity of the antenna system is generally characterized by the System Equivalent Flux Density (SEFD), that can be determined by, 

\begin{equation}
SEFD=\frac{2\times k\times T_{sys}}{A \times \eta \times 10^{-26}},
\label{eq1}
\end{equation}
where k is Boltzmann's constant ($1.38\times10^{-23}~m^2~kgs^{-2}~K^{-1}$), A is the aperture area of antenna ($m^2$), and $\eta$ is the aperture efficiency, which is 0.55 for NTSC antennas. The SEFDs of the three stations are 2647, 2836 and 2457 Jy, respectively. At current stage, the elevation dependency of the SEFD has not been considered yet.

\subsection{VLBI Backend and Data Acquisition System}
Apart from the 13-m antennas and the cryogenic cooling broad-band receiver, each station is equipped with a VLBI digital backend, a hydrogen clock with the stability of $5\times10^{-13}$ @ 1 s and $5\times10^{-16}$ @ 1 day, a weather station and two GNSS receivers (One for geodetic use, the other for time synchronization).

\begin{figure}[h]
\centering
\includegraphics[width=0.8\textwidth]{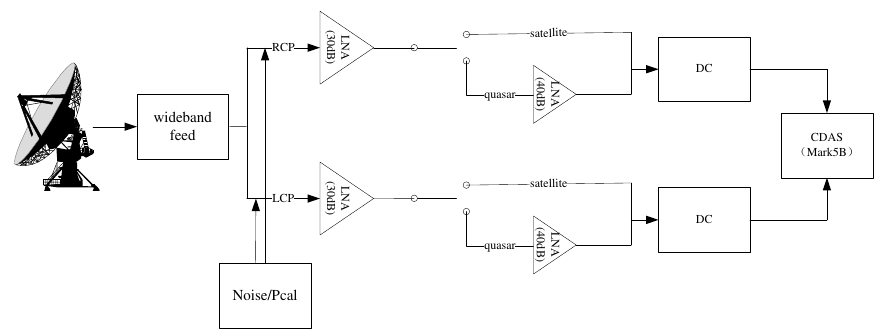}
\caption{Diagram of the signal chain from the feed to the receiver with two LNA to the IF signal outputs from DC to the VLBI recorder.} 
\label{fig5}
\end{figure}

The over-view of the signal chain is shown in Figure~\ref{fig5}. The system temperature calibration (TCAL) and phase calibration (PCAL) block, developed by Shanghai Astronomical Observatory (SHAO), allows for injecting a tone generator of 5 MHz PCAL signal into the signal path at the feedhorn, which is essential to remove instrumental phase differences raised by the receiver and the samplers. The TCAL signal is used to measure the system temperature. As the powers of satellite signals are much stronger than that of radio sources, after the first LNA (30 dB of gain), an optional electronic switch is added to change the signal chain between the quasar and satellite path to avoid saturation of the second LNA (40 dB of gain).

The down-convert (DC, 30 dB of gain) selects broadband signals with local oscillators, phase-locked to a 10 MHz tone from the hydrogen clock, and produces six intermediate frequency (IF) signals: 1) RCP/LCP 50-512 MHz; 2) RCP/LCP 512-1024 MHz; 3) RCP/LCP 50-1024 MHz.

Data digitization and formatting are processed by the Chinese VLBI Data Acquisition System (CDAS, developed by SHAO, \citealt{Zhu+etal+2011}), which is synchronized to the hydrogen clock through 1 pulse-per-second (1PPS) signal. The input signal from DC output is split into two paths in CDAS. One path is to an analog monitoring device that records and displays the channel power on CDAS control software, the other one is to A/D sampler, that digitizes analog signals into 1-bit 50-1024 MHz data streams, or 2-bit 50-512 MHz/512-1024 MHz data streams. The maximal sampling rate is 1024 mega-sample per-second with 2-bit digitization, resulting a 2 Gbps data rate at 512 MHz bandwidth. The CDAS can maximally support $4\times512$ MHz band-width and 8 Gbps data rate (Tabel~\ref{tab3}).

\begin{table}[h]
\begin{center}
\caption{Specifications of VLBI digital backend CDAS}\label{tab3}
\begin{tabular}{ll}
\hline
Parameter& Value \\\hline
Max IF input number& 2 \\
Polarization& RCP/LCP \\
Frequency range& 50-1024 MHz \\ 
&512-1024 MHz \\
&50-1024 MHz \\
Max channel& 32 \\
Each channel band& 32 MHz (2-bit sample) \\
&64 MHz (1-bit sample) \\
Max sampling rate& 2 Gbps \\
Sampling bit& 1/2-bit \\
Max data rate& $4\times 2$ Gbps \\
Out data format& Mark5B \\
\hline
\end{tabular}
\end{center}
\end{table}

After the sampling, the data stream from CDAS is then sent to a record server through fiber-optics, and saved as the Mark5B data format. The record sever is equipped with Redundant Arrays of Independent Drives (RAID) file system, which allows flexible data recording and data transporting. 

\subsection{Operations}
In order to effectively manage the NTSC VLBI array, all three antennas are designed to be remotely controlled via internet. Remote observations can be carried out in the operation room located in Xi'an. Meanwhile, each site has an observing assistant who is responsible to maintain the equipments and carry out on-site observations in case of internet interruption.

The schedule file (VEX file) is produced by the Vienna VLBI and Satellite Software (VieVS, \citealt{Sun+etal+2014}). It can be used in manual mode, where the user explicitly specifies all or part of the information required for schedule, or in automatic mode. For the observations shown in this paper, the automatic mode is used. The VEX file is then transferred to each site via internet. During the observation, the Station Control Software (SCS) on each sites reads in the vex file, then controls and monitors all corresponding on-site devices. Observational information including the azimuths and elevations of antenna, the parameter setup of DC, the CDAS status, etc., is collected by the SCS and saved in a log file for further analysis and fault finding. Before the UT1 observation, usually, a fringe checking on a strong calibrator is carried out to make sure that the system is working properly.

\section{UT1 experiments}\label{sec:3}
\begin{figure}[h]
\centering
\includegraphics[width=0.48\textwidth]{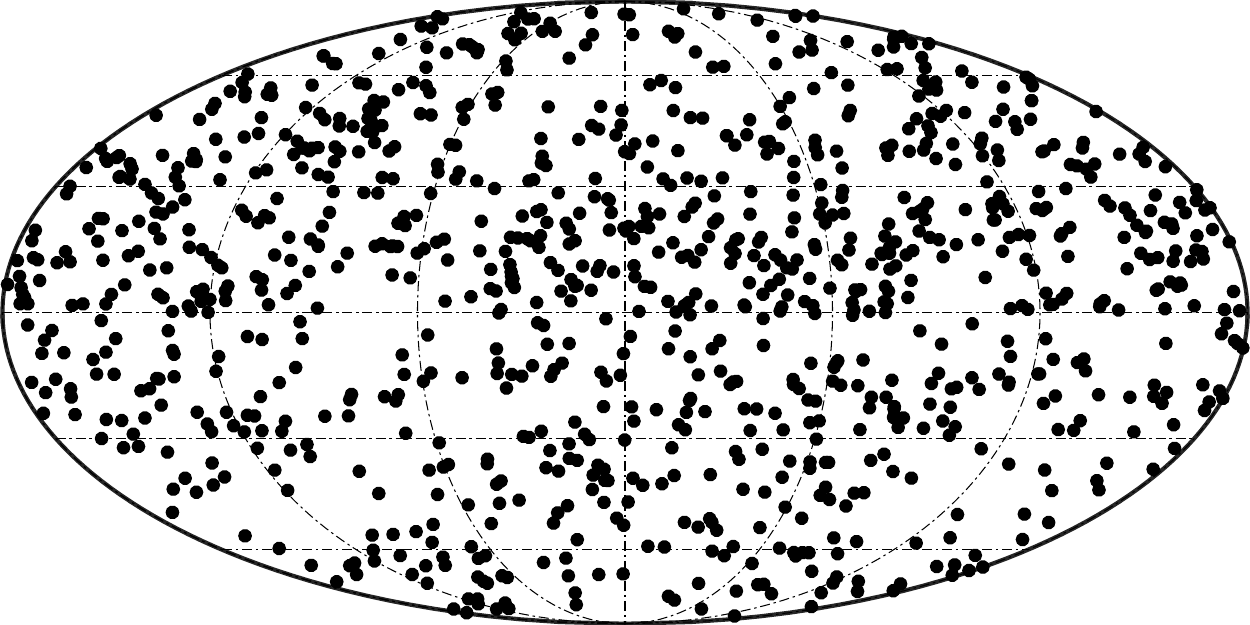}
\includegraphics[width=0.48\textwidth]{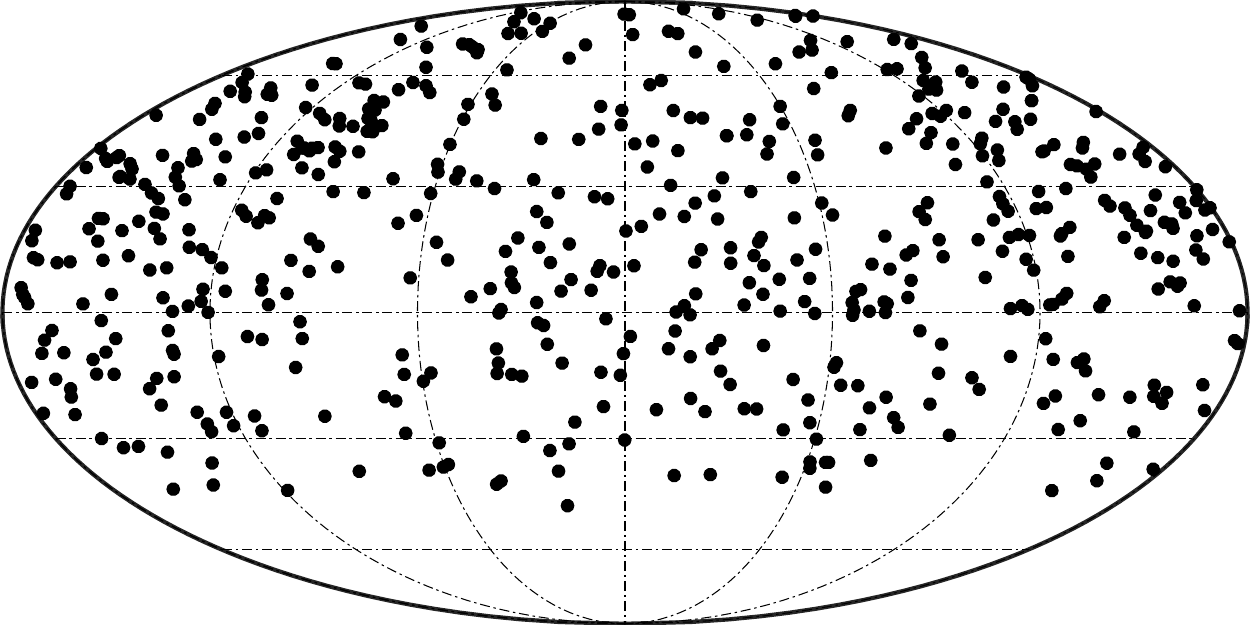}
\caption{The distribution of candidate sources for UT1 experiment. The IVS candidate sources at S/X-band are shown in the left panel. The NTSC candidate sources at C-band are shown in the rigth panel, which are selected from rfc\_2018c.} 
\label{fig6}
\end{figure}

Traditionally, geodetic VLBI experiments are conducted at S-band (2.2-2.3 GHz) and X-band (8.2-8.6 GHz), while, due to the strong RFI at Kashi station, currently, we adopted 4.4 to 4.9 GHz (512 MHz, RCP signal and 2 Gbps) bandwidth for UT1 test experiments. The C-band sources, used for the experiments, are compiled from the rfc\_2018c\_catalogue of compact radio sources\footnote{\it http://alt.astrogeo.org/rfc/} (Figure~\ref{fig6}). The number of sources of our C-band source list is about half of the IVS S/X catalogue.

From June to December of 2018, in total, we carried out 113 UT1 experiments. During July, August, and September, more than one (two to five) sessions were observed within one day. 

With the aim to detect stable fringes, we used several constrains listed below to select sources.
\begin{itemize}
  \item The cut-off elevation angle of $20^\circ$,
  \item The minimum source flux is 0.6 Jy,
  \item The scan duration is limited in 60-200 seconds,
  \item The minimum SNR of 15 at C-band.
\end{itemize}

\begin{figure}[h]
\centering
\includegraphics[width=0.8\textwidth]{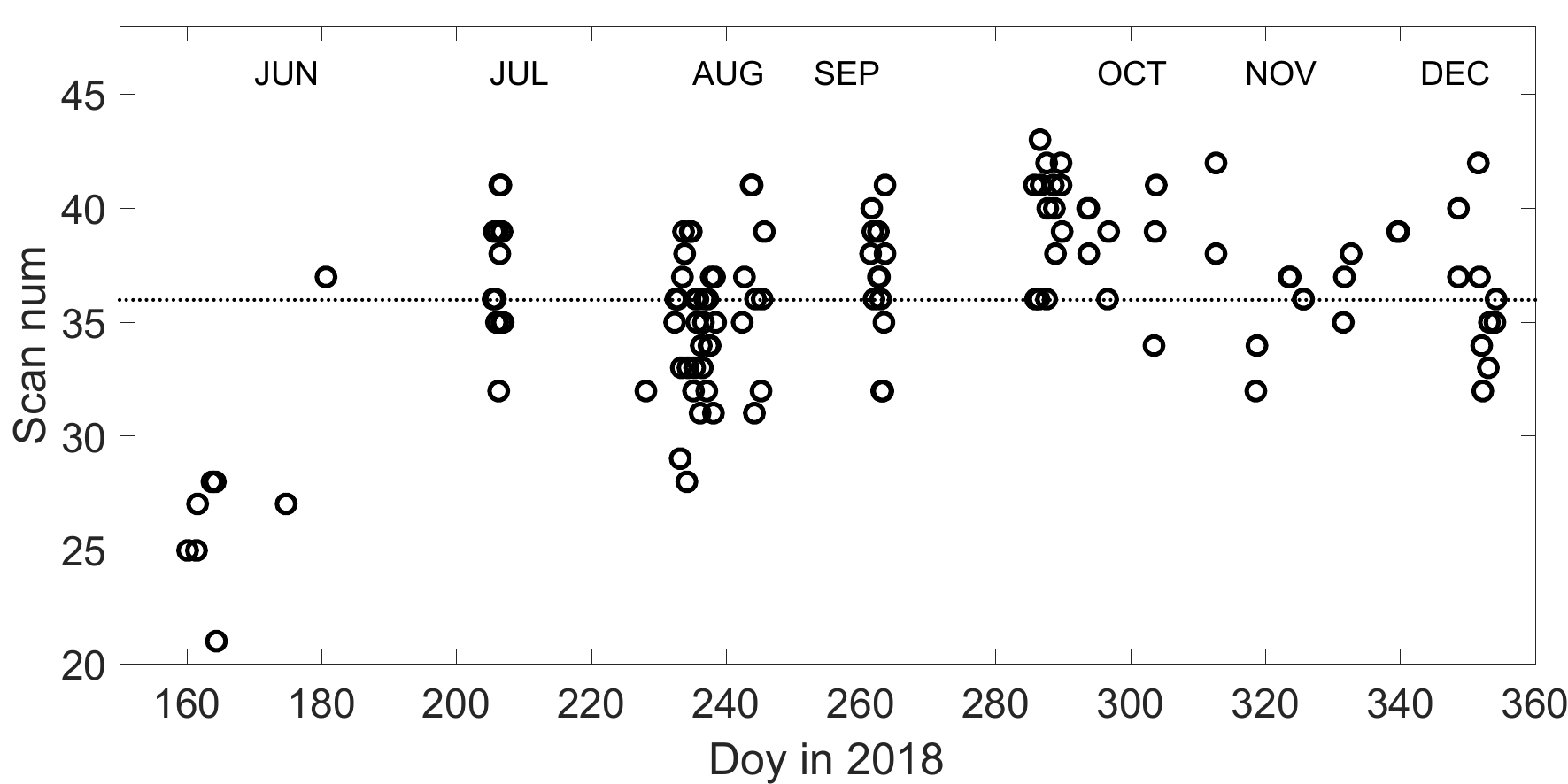}
\caption{The scan number of sessions in 2018. The dot line indicates the average scan number of 36.} 
\label{fig7}
\end{figure}

This results in a mean scan number of 36 for each 1-hour sessions, as shown in Figure~\ref{fig7} and the mean number of sources is 21.

\begin{figure}[h]
\centering
\includegraphics[width=0.8\textwidth]{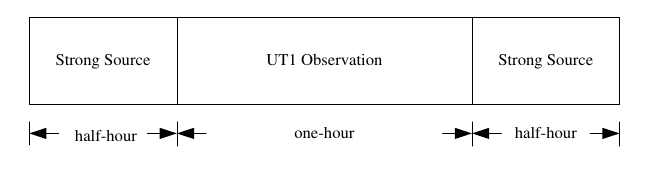}
\caption{Diagram of the 2-hour UT1 sessions.} 
\label{fig8}
\end{figure}

\begin{figure}[h]
\centering
\includegraphics[width=0.8\textwidth]{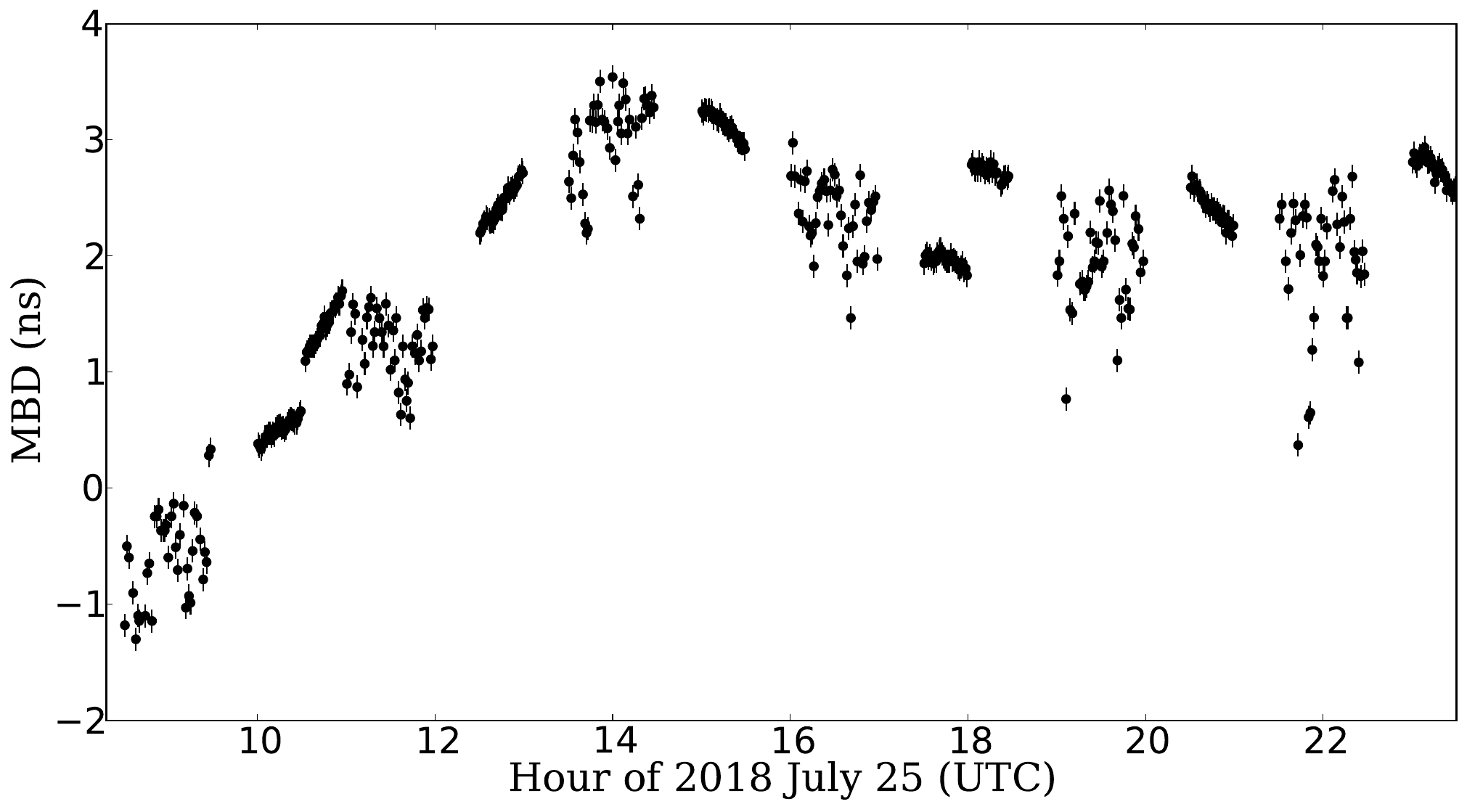}
\caption{The residual multi-band delay (MBD) of 2018 July 25. The dense concentrated points denote the MBD of strong source blocks and the scattered points denote the MBD of UT1 blocks.} 
\label{fig9}
\end{figure}

In addition, we scheduled two half-an-hour strong source blocks before and after the UT1 block (Figure~\ref{fig8}) to monitor the stabilities of hydrogen clocks and to estimate and correct the residual clock drifts. As we found long-term clock and/or instrumental instabilities occasionally. For instance, in Figure~\ref{fig9}, it can be seen a change of clock rate at $\sim$14:00 of 2018 July 25.

\section{Data analysis and results}\label{sec:4}
\subsection{Correlation and post-correlation}
\begin{figure}[h]
\centering
\includegraphics[width=0.8\textwidth]{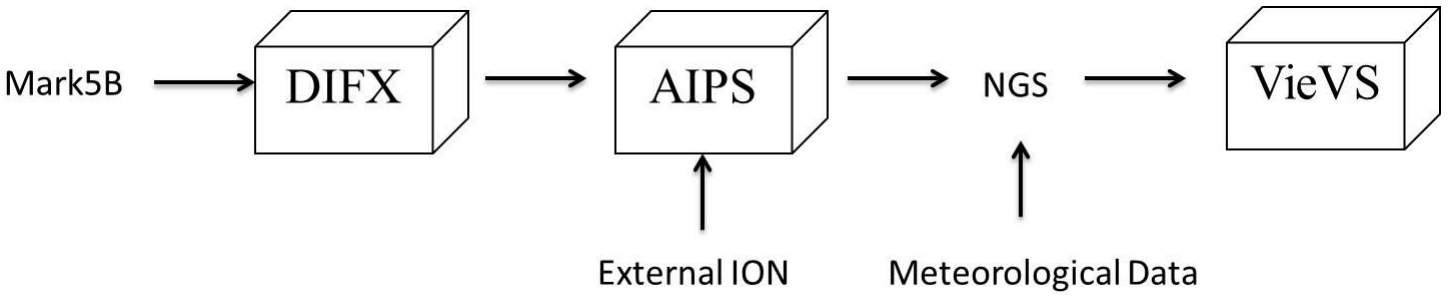}
\caption{Flowchart of the data analysis. The Mark5B data is correlated by the DiFX, the post-correlation is performed by AIPS, finally, the VieVS is used to estimate the parameter.}
\label{fig10}
\end{figure}

The Mark5B raw data from each station are sent to the correlator center by mail for processing, while the small mount (10-30 seconds) of fringe checking data is transferred via internet. The flow diagram of data processing is shown in Figure~\ref{fig10}. The DiFX software (\citealt{Deller+etal+2007}) correlator is used for correlation. It replays the observed data from stations, compensates for the changing phase difference by fringe rotation, then the time-series data is converted into frequency-series data (channelized) prior to cross multiplication. The DiFX software installed in NTSC correlator center is consist of a 20-node cluster with a 200 TB storage space. After correlation, the data are saved as FITS and MK4 format files for further analysis.

The visibility data are processed using a ParselTongue (\citealt{Kettenis+etal+2006}) pipeline script developed by the NTSC VLBI team, the pipeline reduce the FITS file by call the NRAO Astronomical Image Processing System (AIPS) to determine the residual delay and then calculates the total baseline delay: (1) loading and editing of raw uvdata, (2) ionospheric delay calibration, (3) clock drift and manual phase calibration to correct residual equipment delays, (4) bandwidth synthesis to determine the  residual baseline delay, (5) calculating the total ionospheric-free baseline delays.

\begin{figure}[h]
\centering
\includegraphics[width=0.7\textwidth]{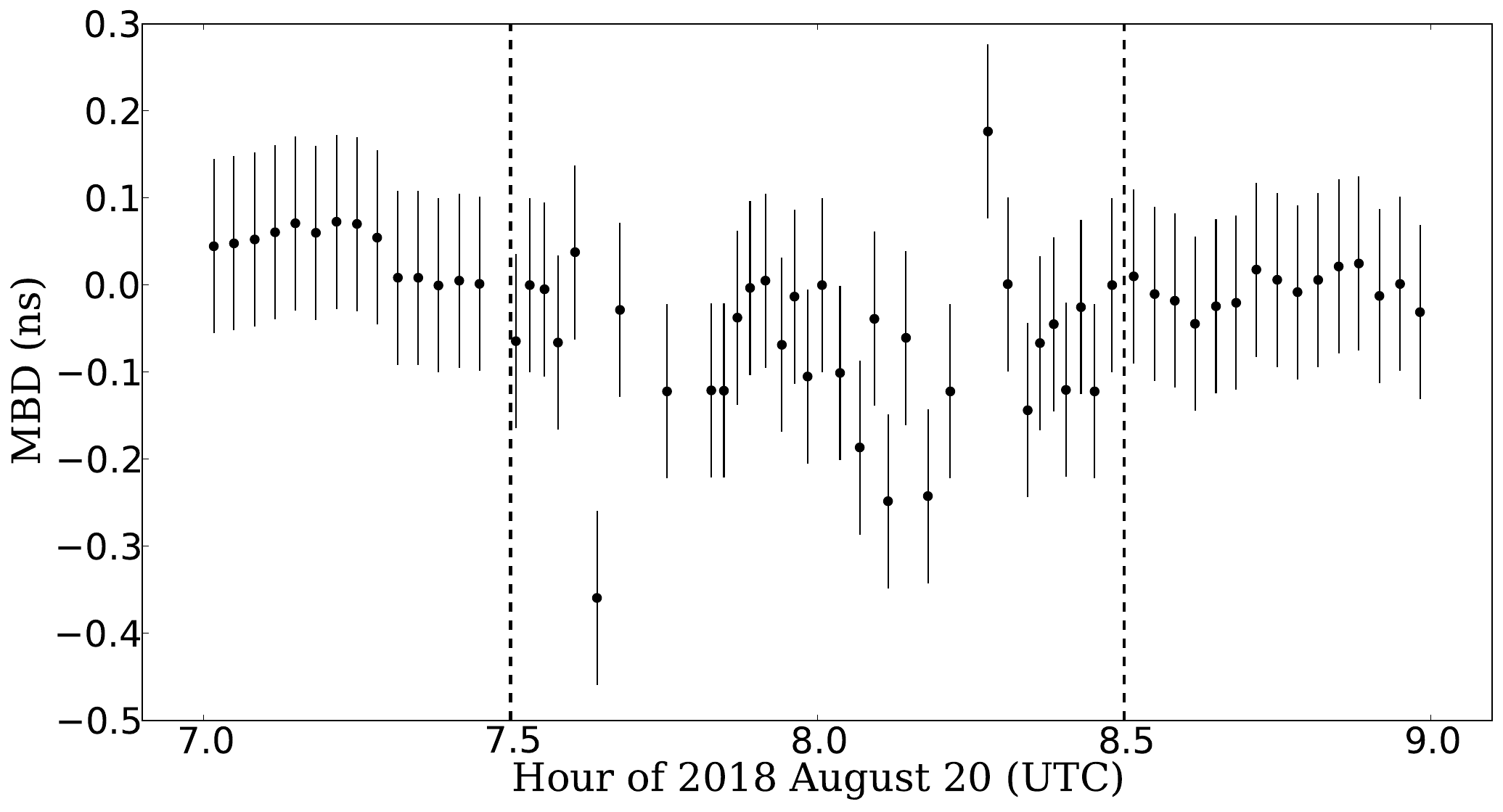}
\caption{The example of the residual MBD resolved by AIPS. The error bar denotes a typical delay uncertainty of 0.1 ns. The two perpendicular dashed lines are the borderline between the UT1 block and strong calibrator blocks.} 
\label{fig11}
\end{figure}

Shown in Figure~\ref{fig11} is an example of the residual Multi-Band Delay (MBD). The residual MBD in most UT1 sessions is in range from -0.2 to 0.2 ns. The two half-hour blocks at the beginning and the end (7.0-7.5 h and 8.5-9.0 h) are the two strong-source blocks, and the scans within 7.5 and 8.5 hours form the 1-hour UT1 block.

\subsection{Estimation of UT1-UTC}
We estimated UT1-UTC by using the VieVS. The observed ionospheric-free total delay from ParselTongue output and the meteorological data (temperature, humidity and atmosphere pressure) collected by the weather stations are combined to generate the ASC II code files developed by US National Geodetic Survey (NGS), which are the inputs for the VieVS.

For propagation delays, we only consider the troposphere influence, as the ionospheric delay is already compensated based on the International GNSS Service (IGS) global model of TEC maps by using the AIPS Task TECOR. The station coordinates were fixed to apriori positions measured by GPS local survey. Regarding station coordinate correction, we adopt the Conventions 2010 provided by International Earth Rotation and Reference Systems Service (IERS) (\citealt{Petit+etal+2010}), that including solid earth tides, tidal ocean loading (FES2004), pole tide and ocean pole tide, while the non-tide atmospheric loading wasn't took into account. For the priori EOP, the IERS C04 14  (here after C04 14) was used and the high frequency (diurnal and subdiurnal) corrections were compensated in polar motion and UT1. The Global Mapping Function (GMF) for tropospheric delay was used for the UT1 test experiment (\citealt{Niko+etal+2015}). Jilin station was chosen as the reference station in the analysis. The detailed parameterization can be seen in Tabele~\ref{tab4}.

\begin{table}[h]
\begin{center}
\caption[]{The analyzed parameters in UT1 determination}\label{tab4}
\begin{tabular}{clcl}
\hline
Parameter& Value &Parameter& Value  \\\hline
Observation type& Group delay  &Precession/Nutation& IAU$^a$ 2006 model\\
Solution type& One standalone &TRP MF& GMF \\
Elevation cut off& $20^\circ$  &dUT1 estimated interval, constraint& 1440 min, 0.0001 \\
Ephemerides& DE 421 ephemeris &ZWD estimated interval, constraint& flexible, 1.5 cm  \\
A priori EOP& EOP C04 14 &Clock estimated interval, constraint& flexible, no const. \\
High frequency EOP& IERS Conventions 2010 &Reference station& Jilin \\ 
Source coordinate& ICRF2 &Station position& ITRF(GPS local survey) \\
\hline
\multicolumn{4}{l}{$^a$International Astronomical Union} \\
\end{tabular}
\end{center}
\end{table}

Then the observation delays and the theoretical delays computed by VieVS were entrance to the least-squares adjustment to determine the best value of the estimated quantities, including the clock offsets, Zenith wet tropospheric delay (ZWD) offsets and UT1 offsets \footnote{The corrections to polar motion and Precession/Nutation weren't estimated due to the limited scan number in one hour observation}.

The 113 UT1 sessions are analyzed by the VieVS, it eliminates outlier delay values, i.e., the scan with residuals larger than 10 cm. Finally, we get  the UT1-UTC estimations for 107 sessions. There are 6 sessions for which we can't estimate the UT1-UTC due to the equipment troubles (e.g., the losing lock of hydrogen clock and breakdown of the record server).

In general, if random noises were the only error source, then the $\chi^2$  of session would be $\sim$1. In reality, we find that the $\chi^2$ of most sessions are in 3-7 in NTSC UT1 experiments, indicating the presence of other error sources and/or modeling errors, for instance, the errors in station coordinates, which require further detailed investigation.

\subsection{UT1 comparison}
\begin{figure}[h]
\centering
\includegraphics[width=0.8\textwidth]{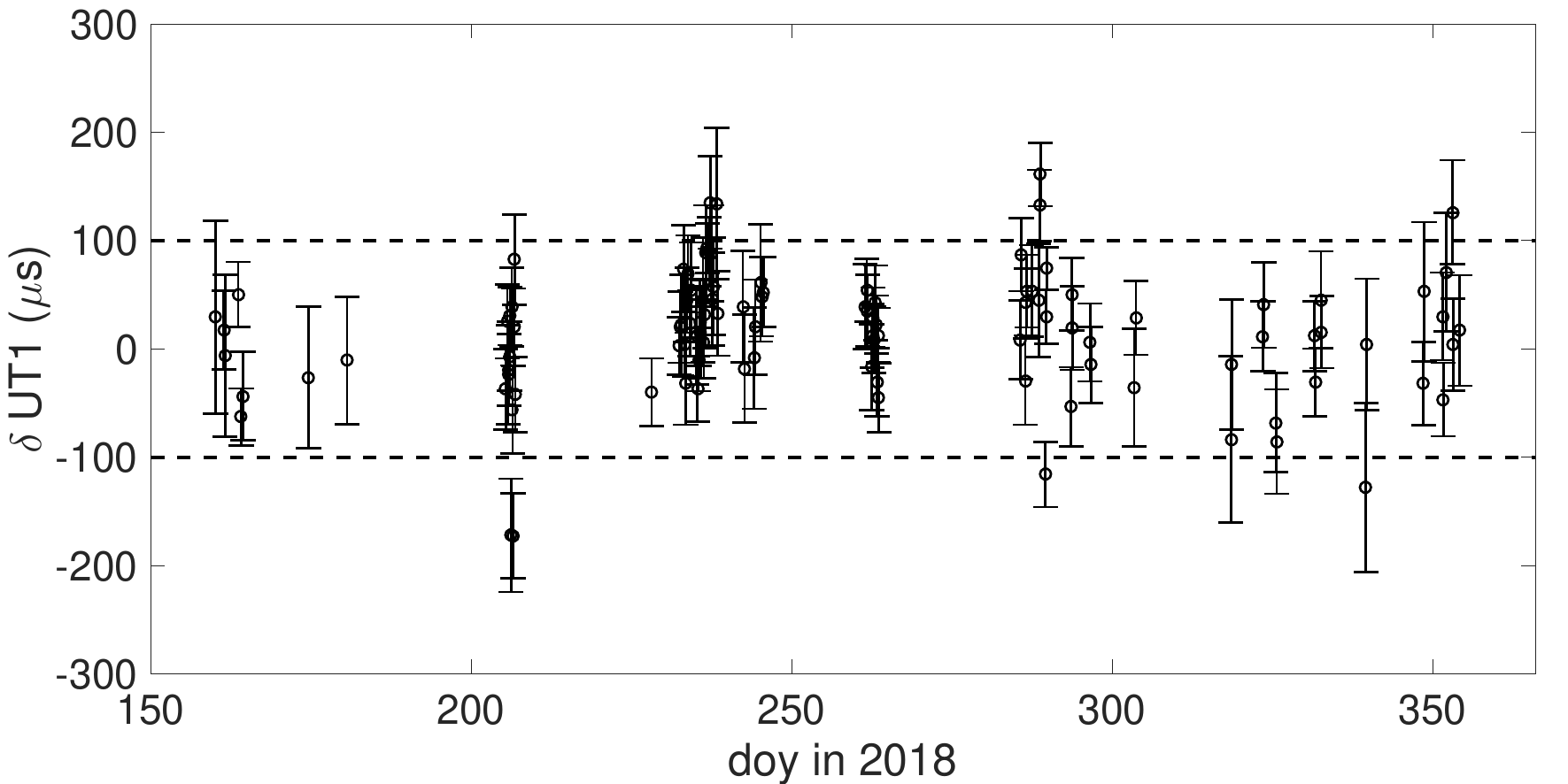}
\caption{Differences between UT1-UTC obtained from NTSC observations and C04 14 series from June to December in 2018. Around 96\% of the points are within the $\pm$100~$\mu$s (denoted by the horizontal dashed lines).} 
\label{fig12}
\end{figure}

\begin{figure}[h]
\centering
\includegraphics[width=0.8\textwidth]{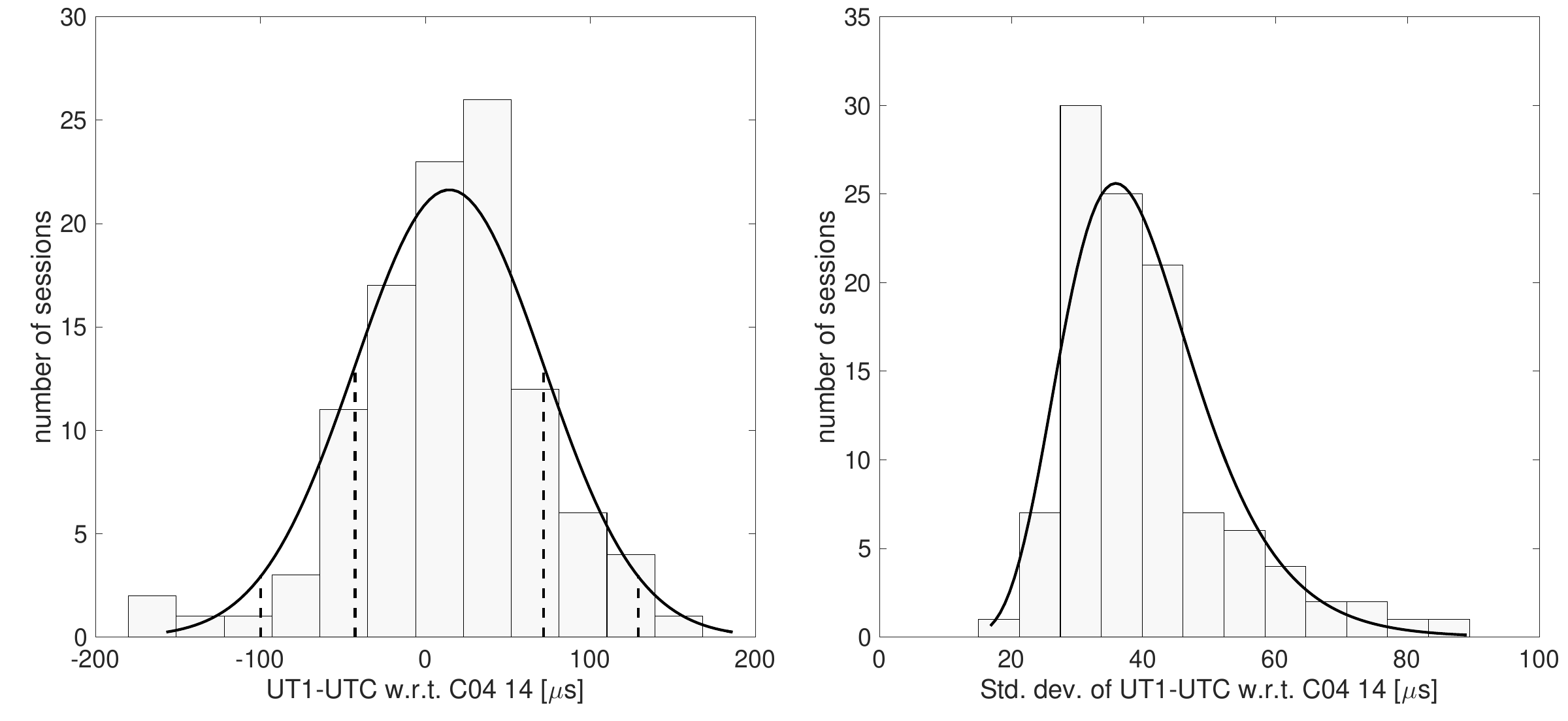}
\caption{Distribution of differences and their formal errors in UT1-UTC between NTSC VLBI and C04 14 from June to December in 2018. The left panel is the distribution UT1-UTC differences w.r.t. C04 14. The right panel is the distribution of the formal errors. The dashed lines in left panel indicate the 1-sigma (-42.7 to 71.7 $\mu$s) and 2-sigma (-99.9 to 128.9 $\mu$s) confidence intervals. The theoretical curves for a Gaussian distribution (left) and an empirical chi-distribution (right) are overlaid.} 
\label{fig13}
\end{figure}

In this subsection, we evaluate the accuracy of UT1 experiments by comparing
our estimation with C04 14 and IVS INT1 results.
Figure~\ref{fig12} shows the differences between our measurement and C04 14. Such difference can be an indicator of the agreement between
our measurement and the true value (accuracy). The histogram of the this
differences is shown in the left panel of Figure~\ref{fig13}, which follows a
Gaussian distribution. On average, there is 14.5 $\mu$s bias between our
estimation and C04 14, and the full width at half maximum (FWHM) is 114.4 $\mu$s
($2\sigma$). 

The 14.5 $\mu$s bias of our results may be partly due to bias of our
station coordinates, and partly due to the station clock offsets, as studied by
\citealt{Hobiger+etal+2009}, that the VLBI reference station clock offsets can result
in a bias of UT1 estimates, and partly due to unkown
absolute electronics delays. As pointed by \citealt{Anderson+etal+2018}, the
full-time offset between the reception of a wavefront by a VLBI antenna, and
the propagation of that wavefront through the receiver electronics, backend
electronics, and associated cables, and the association of a time tag is not
known in VLBI. Further, the absolute offset between VLBI time and Coordinated Universal
Time (UTC) is not known (\citealt{Himwich+2003}, \citealt{Himwich+2005}; \citealt{Himwich+etal+2017};
\citealt{Anderson+etal+2018}). Such bias can only be reduced by further internaiontal
and domestic VLBI sessions. 

The error bars in the Figure~\ref{fig12} are the formal errors of the least-square fitting, which indicates the repeatability of the measurement (precision), the histogram of which is shown in the right panel of Figure~\ref{fig13}. Regarding the repeatability, roughly 82\% of sessions give formal errors for UT1-UTC for less than 40 $\mu$s, following the chi-square distribution.

Further, we compare the accuracy and precision of the NTSC measurement and IVS INT1 result in 2015-2018. In Table~\ref{tab5}, listed are the Root-Mean-Square (RMS), the average values of the difference w.r.t the C04 14 and the formal error. It can be concluded that the UT1-UTC accuracy achieved with the Jilin-Kashi baseline is on average 2-3 times larger than the IVS INT1. Taking into account of the baseline length, the system reliability and the observing frequency, the accuracy currently achieved is reasonable.

\begin{table}[h]
\begin{center}
\caption{The comparison in the UT1-UTC estimation of NTSC and IVS INT1 results w.r.t the C04 14.}\label{tab5}
\begin{tabular}{cccc}
\hline
Type& RMS($\mu$s)& $<\delta$UT1$>$($\mu$s)& $<$formal error$>$($\mu$s) \\\hline
NTSC measured& 58.8& 14.5& 40.3 \\
IVS INT1& 25.5& 6.9& 11.6 \\
\hline
\end{tabular}
\end{center}
\end{table}

\section{Summary and future plan}\label{sec:5}
We have constructed, commissioned and are now operating three new 13-m radio telescopes for geodetic VLBI at Jilin, Kashi and Sanya of China. From June to December, the last half year of 2018, we conducted 113 times UT1 experiments.

In this paper, we presented results of these experiments. By statistics, a bias of 14.5 $\mu$s is found between our estimation and C04 14, and the RMS of UT1-UTC differences w.r.t. C04 14 is 58.8 $\mu$s. The repeatability of the UT1 experiments is about 40 $\mu$s, which is around 3 times larger than the IVS INT1 results. Overall, those results indicate the VLBI network is capable to determine UT1.

Regarding the UT1 uncertainty, the inaccurate station position might be one of the primary error sources, we are going to join the domestic and international 24-hour geodetic VLBI campaigns to improve the position accuracy. In addition, optimized source selection and scheduling would be helpful to reduce the systematic error. We also identified several problems with station equipment, which will be solved in the near future. The serious RFI at the KS station can be overcome by adding a filter to save the observing frequency at 8-9~GHz. Finally, we look forward to make contributions to domestic VLBI measurements of EOP, expecially the UT1.

\begin{acknowledgements}
This work was supported by the National Natural Science Foundation of China (Grant Nos. 11703033, 11673051, 11603001 and U1831136) and the West Light Foundation of the Chinese Academy of Sciences (XAB2016A06). Bo Zhang and Yuan-Wei Wu are supported by the 100 Talents Project of the Chinese Academy of Sciences (CAS). We thank Xiuzhong Zhang, Guangli Wang, Fengchun Su, Renjie Zhu and Shaoguang Guo of the Shanghai observatory (SHAO) for their help in the system construction, thank Wu Jiang of SHAO for the help with the DiFX software. We thank Dr. Minghui XU for his suggestions and his careful reading of this paper. We are grateful to the staffs at JL, SY and KS stations for their operation and maintenance. We greatly acknowledge all the developers of the software packages used in this paper, i.e. DiFX, AIPS and VieVS.
\end{acknowledgements}

\label{lastpage}

\end{document}